\documentclass[aps,prl,twocolumn,showpacs,superscriptaddress,groupedaddress]{revtex4}
\usepackage{graphicx}  
\usepackage{dcolumn}   
\usepackage{bm}        
\usepackage{amssymb}   
\usepackage{slashed}
\usepackage{graphicx}				
\usepackage{amsmath}
\usepackage{mathtools}
\usepackage{tikz}
\usetikzlibrary{arrows,backgrounds}
\usepackage[all]{xy}
\begin{document}

\title{Quantization, Holography and the Universal Coefficient Theorem}
\author{Andrei T. Patrascu}
\address{University College London, Department of Physics and Astronomy, London, WC1E 6BT, UK}
\begin{abstract}
I present a method of performing geometric quantization using cohomology groups extended via coefficient groups of different types. This is possible according to the Universal Coefficient Theorem (UTC).
I also show that by using this method new features of quantum field theory not visible in the previous treatments emerge. The main observation is that the ideas leading to the holographic principle can be interpreted in the context of the universal coefficient theorem from a totally different perspective. 
\end{abstract}
\maketitle
\section{1. Introduction}
It is well known that the number of degrees of freedom associated to the fields of a quantum field theory is infinite [1]. This leads to problems in the description
of quantum field theories involving gravitation and Black Holes [2]. The main approach to these problems until now was based on the holographic assumption [3]. This stated that the number of degrees of freedom associated to a volume in a given space should be mapped unambiguously to a surface surrounding that volume. While the main result of holography is certainly correct (there exists a restriction on the allowed number of degrees of freedom) one can use alternative ideas to derive it. Recently some papers by Gerard 't Hooft [4] proposed new ideas related to the way quantization should be interpreted. While it is my opinion
 that any hidden variable model is inconsistent with available experimental data, these ideas may have another interpretation that does not contradict any aspects of quantum mechanics as understood now. Moreover, they can be interpreted in terms of the Universal Coefficient Theorem (UCT) from algebraic topology [5]. 
The idea behind this theorem is that given a space with certain properties, the information one can extract from that space depends on the accurate choice of the coefficient group used to define the cohomology groups. As cohomology is important in the geometric quantization it may very well happen that properties of the field space of a theory in the presence of a black hole are being misinterpreted while using cohomology groups without or with ill-suited coefficient groups.
As a simple example (see ref. [5]) I may take a Moore space $M(Z_{m},n)$ obtained from $S^{n}$ by attaching a cell $e^{n+1}$ by a map of degree $m$. The quotient map $f:X\rightarrow X/S^{n}=S^{n+1}$ induces trivial homomorphisms on the reduced homology with $Z$ coefficients since the nonzero reduced homology groups of $X$ and $S^{n+1}$ occur in different dimensions. But with $Z_{m}$ coefficients the problem changes, as we can see considering the long exact sequence of the pair $(X,S^{n})$, which contains the segment
\begin{equation}
 0=\tilde{H}_{n+1}(S^{n};Z_{m})\rightarrow \tilde{H}_{n+1}(X;Z_{m})\xrightarrow{f_{*}}\tilde{H}_{n+1}(X/S^{n};Z_{m})
\end{equation}
Exactness requires that $f_{*}$ is injective, hence non-zero since $\tilde{H}_{n+1}(X;Z_{m})$ is $Z_{m}$, the cellular boundary map 
\begin{equation}
H_{n+1}(X^{n+1},X^{n};Z_{m})\rightarrow H_{n}(X^{n},X^{n-1};Z_{m})
\end{equation}
being exactly 
\begin{equation}
Z_{m}\xrightarrow{m}Z_{m}
\end{equation}
One can see that a map $f:X\rightarrow Y$ can have induced maps $f_{*}$ that are trivial for homology with $Z$ coefficients but not so for homology with $Z_{m}$ coefficients for suitably chosen m. This means that homology with $Z_{m}$ coefficients can tell us that $f$ is not homotopic to a constant map, information that would remain invisible if one used only $Z$-coefficients. 
From this simple example we can already see that information is being ``lost'' when thinking in terms of (co)homology groups without coefficients or with naive choices of coefficient groups. Of course this ``loss'' of information should not be taken in any fundamental sense. The information is still there (the space has some topology) but our knowledge about it is blurred by a bad choice of coefficient groups. The idea presented by the UCT may be extended to the field space of the quantum field theories. By adopting this way of looking at the problem one can construct a set of quantum field theories corresponding to the correct description in terms of quantized fields without having to map the problem into a string-type one. This has certain advantages as string field theoretical calculations are notoriously difficult.  
When constructing a QFT using geometric quantization one starts with the introduction of a symplectic manifold associated to a classical problem. This space is then extended according to the Batalin-Vilkovisky prescription. This assures that a symplectic reduced space suitable for quantization always exists. However, one has to keep in mind that there exists a certain gauge freedom associated to the initial action functional. While the BV-BRST prescription introduces a complex equipped with the structure of a differential graded Poisson algebra, it does not tell us how to consider the resulting cohomology groups in order to access all the relevant information available in the field space. In normal situations this lack of precision is not important. However, in the case of black hole physics the proper characterization of the field space appears to be essential.
I show here that introducing a coefficient group one may obtain new information from a field theory in the context of geometric quantization. 
In order to present my results, in chapter 2 I show a short overview of the geometric quantization prescription and all the necessary steps it involves. In chapter 3 I use BV-BRST quantization prescriptions in order to introduce a symplectic space and extend the resulting cohomology groups with a set of integral coefficients. The universal coefficient theorem states that this can always be done without changing the resulting theorems. By this I transfer in some sense 't Hooft's intuition into a mathematically sound formalism without altering standard quantum mechanics and without assuming its breakdown at small distances. In fact quantum mechanics is equally valid in the case of black holes as it is in the case of normal physics. The only caveat is that one has to accurately probe the functional spaces one is trying to describe. In chapter 4 I present how these results relate to the holographic principle and how this interpretation supersedes the holographic idea. In principle an area law can be derived if a particular combination of field space and choice of characterization group is made. However, other choices are also possible, relating theories in different dimensions in a similar way. In fact there exists much arbitrariness in the formulation of "holographic" principles and whole sets of dualities relating various dimensions may emerge. 
I end by concluding that the holographic idea is only one particular interpretation of the universal coefficients theorem of algebraic topology. While intuitively appealing it is not fundamental in the sense of being a "property of reality" but simply the result of a particular choice in the way one regards quantum field theories. 

\section{2. Geometric quantization}

Probably the best mathematical formalization of quantum mechanics is offered by  what is known as ``geometric quantization'' [6]. In this formulation one starts with a classical theory and follows a set of steps that assure the consistency of the resulting quantum theory. According to this principle one may start with a general classical action depending on a set of fields $S[\phi]$. This implies the existence of a symplectic manifold. The main idea is to realize the symplectic form of this manifold as the curvature of a $U(1)$ principal bundle with a connection. We obtain the pre-quantum Hilbert space as the Hilbert space of square integrable sections of the principal line bundle. This Hilbert space is not the one we seek as it overestimates the number of states. In fact one has to pick for each point in the space a certain subspace of the complexified tangent space at that point. One defines the quantum Hilbert space to be the space of all square integrable sections of the line bundle that give $0$ when differentiated covariantly at that point in the direction of any vector of the tangent space. 
As basic quantum mechanics teaches us there exist two sets of variables that become non-commutative operators when quantizing. These may be called ``positions'' and ``momenta'' although their physical meaning may be rather different. 
Independent on the fact that one is working in the simple quantization or, more advanced, in quantum field theory in the context of path integrals, this property remains valid although without being directly seen in the last case. Therefore the next step is the choice of a polarization. This is not unique nor trivial but at this moment it is not necessary to enter in further details. Once a polarization is available one can form a Hilbert space of states as the space of sections of the associated line bundle. The last step would be to associate to the classical variables actual quantum operators on the quantum Hilbert space. This amounts to the quantization of observables while mapping Poisson brackets to commutators. This procedure is in general not well defined for all operators. One can go to the Feynman path-integral formulation. There the information related to the non-commuting operators is encoded in the specific indexation of the c-numbers or Grassmann numbers existing in the theory. One can see that there are several critical steps in this procedure: identifying a symplectic manifold suitable for quantization, the choice of a polarization, the mapping of classical variables to quantum operators while mapping the Poisson brackets to commutators etc. 
I will focus now on the first step. In principle the symplectic phase space suitable for quantization is provided by the BV-BRST formalism. This implies the extension of the usual number of fields by the introduction of the usual ghosts, anti-ghosts and auxiliary fields.
A new structure called anti-bracket emerges in this process. This is defined considering two Grassmann functionals $F$ and $G$ as 

\begin{equation}
(F,G)=\frac{\delta^{r}F}{\delta\phi^{A}(x)}\frac{\delta^{l}G}{\delta\phi^{*}_{A}(x)}-\frac{\delta^{r}F}{\delta\phi^{*}_{A}(x)}\frac{\delta^{l}G}{\delta\phi^{A}(x)}
\end{equation}
involving alternate functional differentiation with respect to the fields $\phi^{A}$ and antifields $\phi_{A}^{*}$. ($r$ and $l$ superscripts meaning right and left derivatives respectively).
The antibracket has some important properties: it changes the statistics as 
\begin{equation}
\epsilon[(F,G)]=\epsilon(F)+\epsilon(G)+1
\end{equation}
and satisfies the following relation
\begin{equation}
(F,G)=-(-1)^{(\epsilon(F)+1)(\epsilon(G)+1)}(G,F)
\end{equation}
where $\epsilon$ is the Grassmann parity operator.
Using this structure the Batalin-Vilkovisky prescription can be written as
\begin{equation}
\frac{1}{2}(W,W)=i\hbar\Delta W
\end{equation}
where 
\begin{equation}
\Delta=(-1)^{\epsilon_{A}+1}\frac{\delta^{r}}{\delta\phi^{A}}\frac{\delta^{r}}{\delta\phi^{*}_{A}}
\end{equation}
W is called the "quantum action" and is a solution of the above equation.

Having the action functional one can define the on-shell region as the space of critical points associated to the action functional 
\begin{equation}
\{\phi \in X | dW[\phi]=0\}
\end{equation}
There exist connected components in the on shell space. These can be identified by the first homotopy group
\begin{equation}
\pi_{0}(\{\phi \in X | dW[\phi]=0\})
\end{equation}
The functions on the classes of this group are the gauge invariant observables. One can observe that the correct identification of possible maps as well as homotopically equivalent structures is extremely important for the correct construction of the field space in the phase preceding actual quantization.
One can define the Batalin-Vilkovisky complex of vector spaces as
\begin{equation}
BV=(0\rightarrow\Gamma(TX)\xrightarrow{dW[*]}C^{\infty}(X)\rightarrow 0)
\end{equation}
where $\Gamma(TX)$ are the sections of the tangent space to the space X and $C^{\infty}$ is the space of infinity differentiable functions.
The cohomology of degree zero of this complex is precisely the space of functions on the critical points (the gauge invariant observables)
\begin{equation}
H^{0}(BV)=C^{\infty}(X)/Im(dW[*])
\end{equation}
Following the BV prescription we have a rule that allows the identification of physical observables. A quantum field theory defined in terms of Feynman's path integrals will involve the integration over the space of inequivalent configurations in order to generate the probability amplitudes. 
When constructing the Lagrangian and defining the action functional one considers all these extensions although in less general situations most of the supplemental fields are integrated out from the very beginning. What is important and represents the novelty in this paper is the way in which one characterizes the cohomology groups associated to the Batalin-Vilkovisky complex. This will be the subject of the following chapter. 
\section{3. (co)Homology with coefficients}
The holographic principle [3] is based on the observation that the number of degrees of freedom accessible in a region of space is widely over-counted by any realistic quantum field theoretical construction. When one considers a region of collapsing matter forming a black hole this over-counting becomes relevant as it interferes with most of the attempts to correct quantization. In essence the holographic principle states that the "information" in a region that collapsed forming a black hole is completely determined by the degrees of freedom on the horizon. The next step would be to restrict the field theories used for the description of this situation to field theories on the surface of the above mentioned boundary. Although this way of thinking has its merits (see the AdS/CFT hypothesis), it appears to me that several aspects have been overlooked. When thinking in this way one assumes that the field theoretical construction is unique in the following sense: once the cohomology groups of the BV-complex are established there is no freedom in the way one analyses them. However, in reality there may appear topological features that are invisible unless a specific choice of coefficients is made. While the new features are always in the quantum field theory (encoded in the topology of the field space) our access to it is limited by a poor choice of the coefficient group. This is to say that quantum field theory in principle must not be replaced with other "more fundamental" theories (see string theory) in order to provide the full information about reality. It merely needs to be treated according to the specificities of each situation and in a mathematically well defined way. By doing this, maps considered to be trivial may appear to be relevant and to generate symmetries in the field space that were not observed before. In order to explain these ideas I will start with a description of gravitation in the context of the Einstein-Hilbert action functional
\begin{equation}
S=\frac{1}{2k}\int R\sqrt{-g} d [vol_{M}]
\end{equation}
where 
\begin{equation}
g=det(g_{\mu\nu})
\end{equation}
$R$ is the Ricci scalar, $g_{\mu\nu}$ is the space-time metric, $k=8\pi G c^{-4}$, $G$ being the gravitational constant, $c$ the speed of light in vacuum and the configuration space $\Sigma(M)=(T^{*}M)^{2\otimes}=T^{0}_{2}M$ is a space of rank $(0,2)$ tensors.
It is well known that general relativity is diffeomorphic invariant. The main associated problem is the definition of diffeomorphism invariant observables. A systematic characterization of these can be an important step towards quantization. In order to apply the ideas presented in the previous chapters I need a definition of the BV-complex for general relativity. I follow for this construction reference [6] in some extent but I skip the more involved computational details due to space limitations. In principle fields in general relativity can be treated as natural transformations [7] between functors associated to the configuration space $\Sigma(M)$ and functors associated to the space of the functionals of a given action $S[\Phi]$. Denote the space of natural transformations of this form as $Nat(\Sigma,C\Sigma)$. Note that the action of infinitesimal symmetries on the elements of $Nat(\Sigma,C\Sigma)$ has two terms, the first being the analogue of the infinitesimal transformations in gauge theories while the second term being characteristic to the theories where symmetries are a consequence of diffeomorphism invariance. One can introduce a Chevalley-Eilenberg complex on the natural transformations defined as 
\begin{equation}
\bigoplus_{k}^{\infty}Nat(\Sigma^{k},C\Sigma)
\end{equation}
where $C\Sigma$ is the graded algebra of the smooth well behaved maps with associated well behaved differential
\begin{equation}
\delta:\bigwedge^{q}\chi(M)\otimes F(M)\rightarrow\bigwedge^{q+1}\chi(M)\otimes F(M)
\end{equation}
where $\chi(M)=\Gamma[TM]$ is the set of sections of the tangent manifold, with the action:
\begin{widetext}
\begin{equation}
(\delta\omega)(\xi_{0},...\xi_{q})=\sum_{i=0}^{q}(-1)^{i}\partial_{\rho M(\xi)}(\omega(\xi_{0},...\xi_{q}))+\sum_{i<j}(-1)^{i+j}\omega([\xi_{i},\xi_{j}],...,\hat{\xi}_{i},...,\hat{\xi}_{j},...,\xi_{q})
\end{equation}
\end{widetext}
The BV-differential can be defined analogously. Skipping some mathematical details related to the definition of the necessary structures [6] the definition is:
\begin{equation}
(\delta_{BV}\Phi)_{M}(*)=\{\Phi,L\}_{M}(*,1)+(-1)^{\epsilon(\Phi)}\Phi_{M}(\mathcal{L_{(.)}}*)
\end{equation}
where $\Phi\in Nat(\Sigma,F)$, $L$ is the corresponding Lagrangian, $\mathcal{L}$ is the Lie derivative, * replaces the function where the evaluation is performed and . represents the direction of the Lie derivative. (*,1) represents the evaluation of the function on the support where its distribution is $1$, i.e. where it is well defined. 
It is certain that the 0-cohomology of $\delta_{BV}$ is not trivial as it contains for example the Riemann tensor contracted to itself
\begin{equation}
\Phi_{(M)}=\int_{M}R_{\mu\nu\alpha\beta}R^{\mu\nu\alpha\beta}d[vol_{M}]
\end{equation}
The physical quantities should be identified with the elements of the cohomology of the BV-differential $H^{0}[\delta_{BV}]$.
Of course the next step would be to introduce a Poisson bracket and to obtain gauge fixed classical Lagrangians. 
Indeed this can be done. 
\begin{equation}
L^{g}=L^{EH}+L^{FP}+L^{GF}
\end{equation}
this being the standard "gauge fixed" lagrangean with $L^{EH}$ being the original Einstein-Hilbert action, $L^{FP}$ being the "Faddeev-Popov" term and $L^{GF}$ the gauge fixing term.
One can write these terms down using several auxiliary fields and ghosts
\begin{equation}
L_{M}^{EH}=\int d[Vol_{M}]\sqrt{-g}R
\end{equation}
\begin{equation}
\begin{array}{l}
L_{M}^{FP}=i\int d[Vol_{M}]\frac{1}{\sqrt{-g}}\nabla_{\nu}\bar{C}_{\mu}(g^{\lambda\nu}\nabla_{\lambda}C^{\mu}+\\
+g^{\lambda\mu}\nabla_{\lambda}C^{\nu}-\nabla_{\lambda}C^{\nu}-\nabla_{\lambda}(C^{\lambda}g^{\mu\nu}))\\
\end{array}
\end{equation}
\begin{equation}
L_{M}^{GF}=-\int d[Vol_{M}](\frac{\alpha}{2}B^{\mu}B_{\mu}+\frac{1}{\sqrt{-g}}B_{\mu}\nabla_{\nu}g^{\mu\nu})
\end{equation}
where $C^{\mu}$ are ghost fields and $B^{\mu}$ are auxiliary fields. The explicit construction of these terms as well as the transformation laws for the new fields are not of major concern in this work and can be cross-verified in reference [6].
The quantization prescription states that after doing this an integration involving inequivalent space-time configurations has to be constructed. This integral is naively ill-defined. However one may observe that in the previous steps no reference to the coefficients of the cohomology group has been made. In fact why would one do that? The cohomology groups are isomorphic anyway. This is true only when the $Tor$-group playing the role of an obstruction to the isomorphism is trivial. The explicit calculation and the integration over inequivalent classes implies however the correct characterization of the properties of the space in any situation. This doesn't happen independent of the choice of the coefficient groups. It appears in this way that renormalizability of a quantum field theory of gravity may require not that much an underlying structure (strings) but more a method of correctly identifying the cohomology classes.  As shown in the first example a naive integration may overlook topological features in the field space. These may form equivalence classes on their own while remaining unobserved in any of the current approaches and this will on its turn falsify the identification of physical observables as functions over homotopy groups. In this sense renormalization would require a very special form of regularization. I call this "topological regularization" in order to make the connection with the general idea of regularization of a theory meaning originally to identify the sources of divergence. This is not directly addressed by imposing an underlying structure (string theory or string field theory) but by observing that some unintended and arbitrary choices made during the process of quantization are in fact of a relative (subjective) nature.
One example would be the choice of the coefficient group $Z_{2}$ which I pick arbitrarily now. As this group is used in a standard way for proving the Borsuk-Ulam theorem [5] I will try an extension of this theorem to the present case. I am doing this only in order to give a simple example of the importance of coefficient groups in the quantization of field theories. Many other applications are possible, especially for $Z_{p}$ where $p$ is a prime number. I do not claim that this special example is of major relevance to the quantization of gravity although it may become a model for more important observations. Take a functional 
\begin{equation}
F[*] : S^{n}\rightarrow R^{n}
\end{equation}
and suppose you are integrating over the field space in order to construct the quantization. If you look at the cohomology group with integral coefficients and suppose 
\begin{equation}
F[\Phi]\neq F[-\Phi]\: \forall\Phi\in Nat(\Sigma,C\Sigma)
\end{equation}
 then the functional $G[\Phi]=F[\Phi]-F[-\Phi]$ is never equal to $0$. From this the functional 
\begin{equation}
 H: S^{n}\rightarrow S^{n-1}
 \end{equation}
 given as 
 \begin{equation}
 H[\Phi]=G[\Phi]/||G[\Phi]||
 \end{equation}
  is constant. However $H[-\Phi]=-H[\Phi]$. Therefore the restriction of $H$ to $S^{n-1}$ given by the functional 
  \begin{equation}
  K[\Phi]=H|S^{n-1}:S^{n-1}\rightarrow S^{n-1}
  \end{equation}
   is of odd degree and induces a non-trivial map in the cohomology group $H^{n-1}(S^{n-1})$ although it factors through $H^{n-1}(S^{n})=0$. This is a contradiction which makes the statement that $F[\Phi]\neq F[-\Phi]$ false. By that one observes that the field integration may result in over-counting of field structures when integrating over cohomology groups with undefined coefficients. 
The effect on the above mentioned "pseudo-gauge-fixed" lagrangean is that when performing the quantization the BV-formalism as described there is not sufficient unless one choses the suitable structures for the coefficient groups. Moreover, when dealing with infinite dimensional spaces even this simple identity may become problematic. 
\section{4. Holographic Principle}
As stated already before, the holographic principle is based on the fact that there must exist a map between a volume of normal space and a surface encapsulating a spherically-homotopic volume. This idea leads to the observation that the number of degrees of freedom associated to a QFT defined on the normal volume of space is far larger than the supposed natural degrees of freedom to be associated to a surface area. While this is certainly true, this observation must not be generalized or considered fundamental. As one can see, the problem lies not in the fact that there exists a mysterious way in which the information must be encoded on an area but simply in ignoring some topological features of the field-space one integrates when quantizing a theory. A space appearing as structureless when analyzed using some cohomology groups (associated to the BV-complex) in fact presents much more structure if analyzed using a different coefficient structure. All this structure is not taken into account when performing a field-space integral which obviously leads to divergencies and may lead one to believe the nature is "holographical". 
\section{5. Conclusion}
As a conclusion the interplay between various mappings on functional spaces and (co)homology groups has an important effect on the prescriptions of quantization. This results in a relativization of the holographic principle. One can in principle derive various other forms of "holography" while playing with the methods of probing topological spaces. As one may see now, the holographic principle appears as a necessity only when specific arbitrary choices in the quantization prescription are being made. Hence holography depends on the \textit{method} of quantization and not on quantization itself. Because of this, various choices may alter the form of the principle and even make it \textit{unnecessary} without altering the physical reality. This paper invites to caution when using the holographic principle as a means for proving various conjectures outside its scope. 
Further generalizations to the (co)homologies with local coefficients may also be important when considering quantum field theories of gravity. 

\end{document}